\documentclass[twocolumn,aps,prl]{revtex4-1}

	\usepackage[usenames,dvipsnames]{xcolor}
	\usepackage{amsmath}
	\usepackage{amsfonts}
	\usepackage{amssymb}
	\usepackage[colorlinks=true,citecolor=blue,linkcolor=red]{hyperref}
	\usepackage{graphicx}
	\usepackage{bbold}					
	\usepackage[makeroom]{cancel}		
	\usepackage{multirow}				
	\usepackage[normalem]{ulem}        
	\usepackage{array}

	\newcolumntype{x}[1]{>{\centering\let\newline\\\arraybackslash\hspace{0pt}}p{#1}}

	\DeclareMathAlphabet{\mathbbold}{U}{bbold}{m}{n}

	\newcounter{subeqn} %
	\makeatletter
	\@addtoreset{subeqn}{equation}
	\makeatother

\definecolor{TB}{rgb}{1,0.5,0}
\definecolor{TB}{rgb}{0,0,0} 

\definecolor{ZX}{rgb}{0.4,0,1}
\definecolor{ZX}{rgb}{0,0,0}

\begin{document}
\title{Electron Pulse Compression with Optical Beat Note}
\author{Zhexin Zhao$^1$}
\author{Kenneth J. Leedle$^1$}
\author{Dylan S. Black$^1$}
\author{Olav Solgaard$^1$}
\author{Robert L. Byer$^1$}
\author{Shanhui Fan$^1$}
\affiliation{$^1$Ginzton Laboratory, Stanford University, Stanford, CA 94305, USA}
\date{\today}

\begin{abstract}
Compressing electron pulses is important in many applications of electron beam systems. 
In this study, we propose to use optical beat notes to compress electron pulses. The beat frequency is chosen to match the initial electron pulse duration, which enables the compression of electron pulses with a wide range of durations. This functionality extends the optical control of electron beams, which is important in compact electron beam systems such as dielectric laser accelerators. We also find that the dominant frequency of the electron charge density changes continuously along its drift trajectory, which may open up new opportunities in coherent interaction between free electrons and quantum or classical systems.
\end{abstract}
\maketitle

\emph{Introduction --}
Measurement of fundamental physics on ultrafast timescales is a cornerstone of modern physics, materials science, and biology. Measuring interactions on femtosecond (fs) timescales requires fs probes \cite{kealhofer2016all, morimoto2018diffraction}. Compression of electron beams, which are probes \textit{par excellence} for matter due to their large scattering cross section and high spatial resolution, is therefore of paramount importance to fields such as Ultrafast Electron Diffraction \cite{zewail20064d, baum2006breaking}. 
However, due to the intrinsic time uncertainty of the electron emission process and the errors in the electron gun, the electron pulse generated by the fs photo-emission laser pulse is typically picoseconds (ps) in duration. 
Terahertz (THz) electron compressors have been demonstrated which compress electron pulses from ps to fs timescales \cite{kealhofer2016all, zhang2018segmented, ehberger2019terahertz, zhao2020femtosecond}. 
To compress the whole electron pulse to a single density peak, the periodicity of the electromagnetic wave should be larger than the electron pulse duration. 
In order to compress the whole ps electron pulse, THz electromagnetic fields are therefore used. However, the generation efficiency of THz waves is typically low \cite{huang2013high}.
Compressing electron pulses using optical waves at infrared wavelength range is attractive due to wide availability and high energy efficiencies of such sources. Recent works have demonstrated that the optical near-field can modulate the free electrons to form a train of micro-bunches of sub-fs duration \cite{morimoto2018diffraction, kozak2019all, kozak2018ponderomotive, black2019net, schonenberger2019generation, feist2015quantum, priebe2017attosecond, polman2019electron, rivera2020light, morimoto2020single}.
These demonstrations compress the electrons that fall within a fs optical cycle to a density peak. However, there have not been any previous works indicating the possibility of using optical modulation with fs periodicity to compress the whole ps electron pulse. 

\begin{figure}
    \centering
    \includegraphics[width=0.98\linewidth]{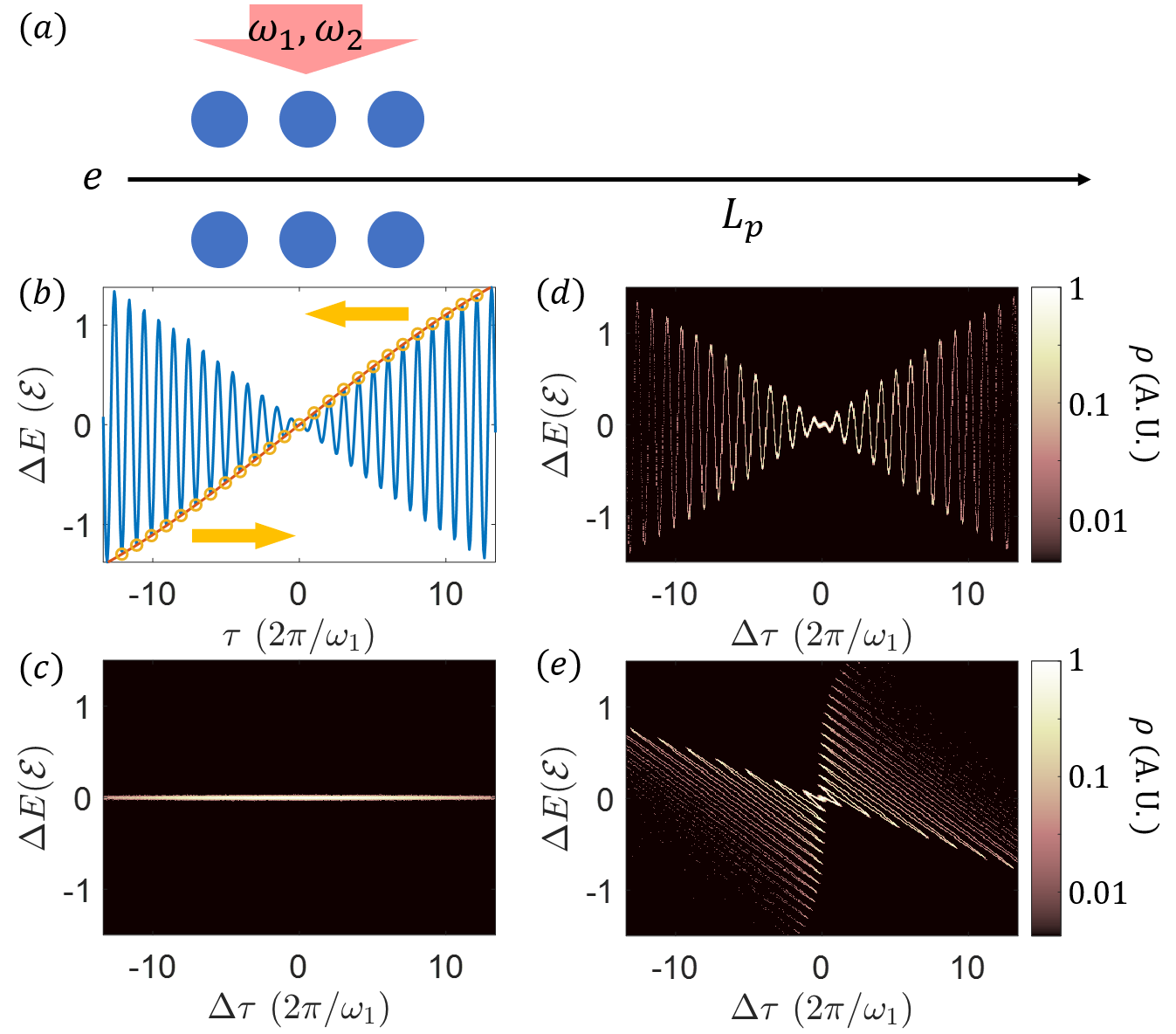}
    \caption{(a) Schematic of the optical electron pulse compressor. The pink arrow illustrates the driving optical pulses with central frequencies $\omega_1$ and $\omega_2$. The blue circles represent the nano-structure, e.g., DLAs, which provides the near-field to modulate the energy of the electrons. The electron travels along the central channel. The drift distance after the energy modulation is $L_p$. (b) Energy modulation near the optical beat node (blue). The orange circles represent clusters that drift to form the density peak, whose drift trajectories are indicated by the orange arrows. (c),  (d), and (e) are initial, immediately after modulation, and final electron distribution in energy-time phase space respectively. For illustration, the electron pulse is 0.1 ps, $1/10$ the duration used in Figs.\ \ref{fig:2freq_2h}-\ref{fig:fourier}.}
    \label{fig:schematic}
\end{figure}

In this Letter, we propose an electron compressor using the optical beat note generated by two infrared lasers to achieve effective THz compression without direct THz generation. If the center of the electron pulse overlaps with the node of the beat pattern, the electron pulse can be compressed after two-color energy modulation and a particular free-space drift (Fig.\ \ref{fig:schematic}). 
For instance, we numerically demonstrate the compression of the full width at half maximum (FWHM) of the electron pulse from 1 ps to $<$ 3 fs. The FWHM indicates the fastest temporal dynamics the electron pulse can probe. Such short FWHM is crucial in time-resolved electron diffraction and microscopy, streak cameras, and dielectric laser accelerators (DLAs) \cite{kealhofer2016all, morimoto2018diffraction, black2019net, schonenberger2019generation}.
To compress electron pulses with different initial durations, one can tune the frequency difference of the two lasers. 
This mechanism allows one to compress electrons with a wide range of initial durations.
The proposed compressor bypasses the difficulty associated with the low efficiency in generating THz fields from infrared lasers with optical nonlinearity, by instead utilizing the nonlinear response of the electron beam to external electromagnetic field. In such response, the difference frequency of the two lasers can play a dominant role.
Furthermore, the dominant Fourier component of the electron density distribution changes along the drift trajectory, providing a tunable resonant frequency when interacting with classical or quantum systems \cite{gover2020free, karnieli2021coherence, kfir2021optical, zhao2021quantum, ruimy2021toward}. 
Finally, this optical electron pulse compression can be implemented in many light-electron interaction systems, such as DLAs \cite{england2014dielectric, breuer2013laser, peralta2013demonstration, cesar2018high, cesar2018enhanced, mcneur2018elements, shiloh2021particle, leedle2015dielectric, leedle2018phase, black2019laser, black2019net, schonenberger2019generation, black2020operating}, which utilize the optical near field to interact with free electrons and can provide large energy modulation.

\emph{Principles --} The schematic of the optical electron pulse compression system is shown in Fig.\ \ref{fig:schematic}(a). The electron pulse is energy modulated in the vicinity of a nano-structure, which is excited by incident optical fields with two frequencies, and drifts in the free space after the modulation. We take the nano-structure as a DLA which can provide around 1 keV energy modulation within 10 $\mu$m interaction distance. Such a large energy modulation is useful in electron compression. 
When the DLA is illuminated by an unchirped optical pulse with central frequency $\omega$, the $z$-component of the electric field at the electron channel takes this form $E_z(z,t) = \Re[E_\omega(z) \exp(-i\omega t)]A(t)$,
where $E_\omega(z)$ is the complex spatial distribution of the field, and $A(t)$ describes the slow-varying pulse envelope of the pulse with unity peak value. In this study, we take the envelope as a Gaussian function, i.e. $A(t) = \exp(-t^2/2\sigma_p^2)$ with $\sigma_p = \Delta_p/2\sqrt{2\textrm{ln}2}$, where $\Delta_p$ is the FWHM of the optical pulse. 
By controlling the driving laser, the transverse deflection forces at the channel center can be cancelled \cite{leedle2018phase, black2020operating, zhao2020mimosa}. Thus, we concentrate on the longitudinal phase space. 
Suppose that the center of the modulation stage is at $z = 0$. If an electron with velocity $v$ passes $z = 0$ at time $\tau$, its trajectory is $z = v(t - \tau)$. The energy modulation on this electron is $\Delta E(\tau) \approx \Re[e\int dz E_\omega (z) \exp(-i\omega z/v) \exp(-i\omega\tau)]A(\tau)$, where we make the approximation that the envelope does not change while the electron traverses the modulation stage.
Taking $e\int dz E_\omega (z) \exp(-i\omega z/v) = \mathcal{E}\exp(i\phi)$, where $\mathcal{E}$ and $\phi$ represent the magnitude and phase of the energy modulation respectively, we find that the electron energy modulation is $\Delta E(\tau) = \mathcal{E}A(\tau)\cos(\omega\tau - \phi)$.

\begin{figure}
    \centering
    \includegraphics[width=0.86\linewidth]{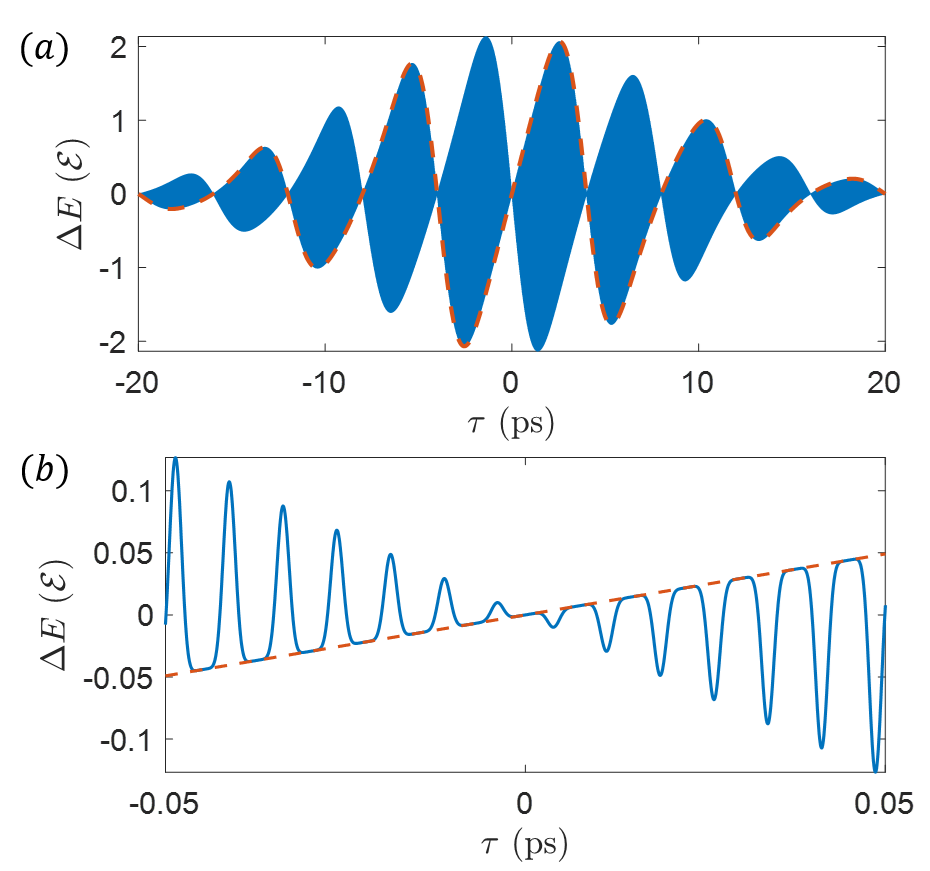}
    \caption{(a) Energy modulation with the first 4 harmonics of two driving optical pulses (blue), given by Eq.\ (\ref{eq:DeltaE_Nh}) with $N=4$. The dashed red curve is the slow-varying $\Delta\bar{E}(\tau)$. (b) Zoom-in view for $\tau$ near zero. The two optical pulses have fundamental frequencies centered at 133.84 THz and 133.59 THz respectively, with 20 ps pulse duration.}
    \label{fig:2freq_2h}
\end{figure}

\begin{figure}
    \centering
    \includegraphics[width=0.86\linewidth]{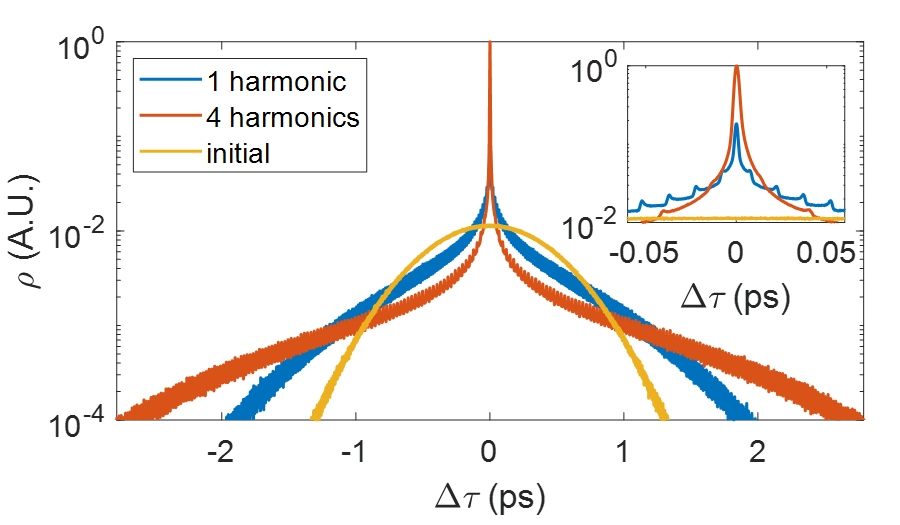}
    \caption{Electron density distribution before (yellow) and after the compression with only fundamental frequencies (blue) or first 4 harmonics (red). $\Delta\tau$ is the time difference between an electron and the electron pulse center passing an observation point. The fundamental frequencies of the two optical pulses are the same as in Fig.\ \ref{fig:2freq_2h} with modulation amplitude $\mathcal{E}=400$ eV. The electron pulse has initial kinetic energy 57 keV with 1 eV uncertainty and 1 ps duration. For normalization, the time integrated densities are kept the same. The inset shows a zoom-in view of the peak.}
    \label{fig:case_study}
\end{figure}

\begin{figure}[th]
    \centering
    \includegraphics[width=0.92\linewidth]{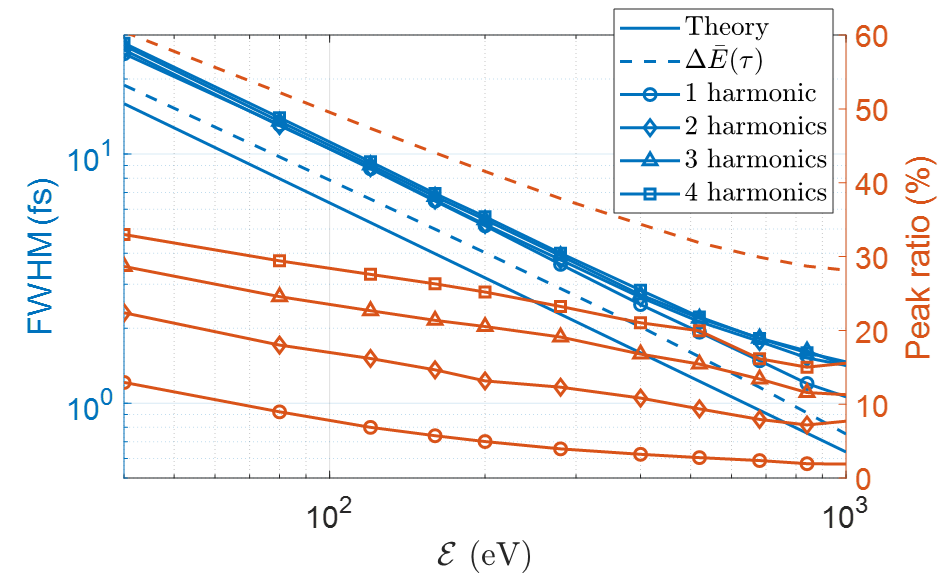}
    \caption{FWHM (blue) and peak ratio (red) of the compressed electron pulse as a function of the energy modulation amplitude. The parameters of the optical pulses and the electron pulse are described in the caption of Fig.\ \ref{fig:2freq_2h} and Fig.\ \ref{fig:case_study} respectively. Solid line is the theoretical FWHM $\delta E_0/\mathcal{E}(\omega_1-\omega_2)$. Dashed line represents the modulation with the slow-varying modulation $\Delta\bar{E}(\tau)$ (Eq.\ (\ref{eq:DeltabarE_Nh}), N=1). The circle, diamond, triangle and square represent respectively the compression with the first 1, 2, 3, and 4 harmonic(s).}
    \label{fig:fwhm_ratio}
\end{figure}

\begin{figure}
    \centering
    \includegraphics[width=0.96\linewidth]{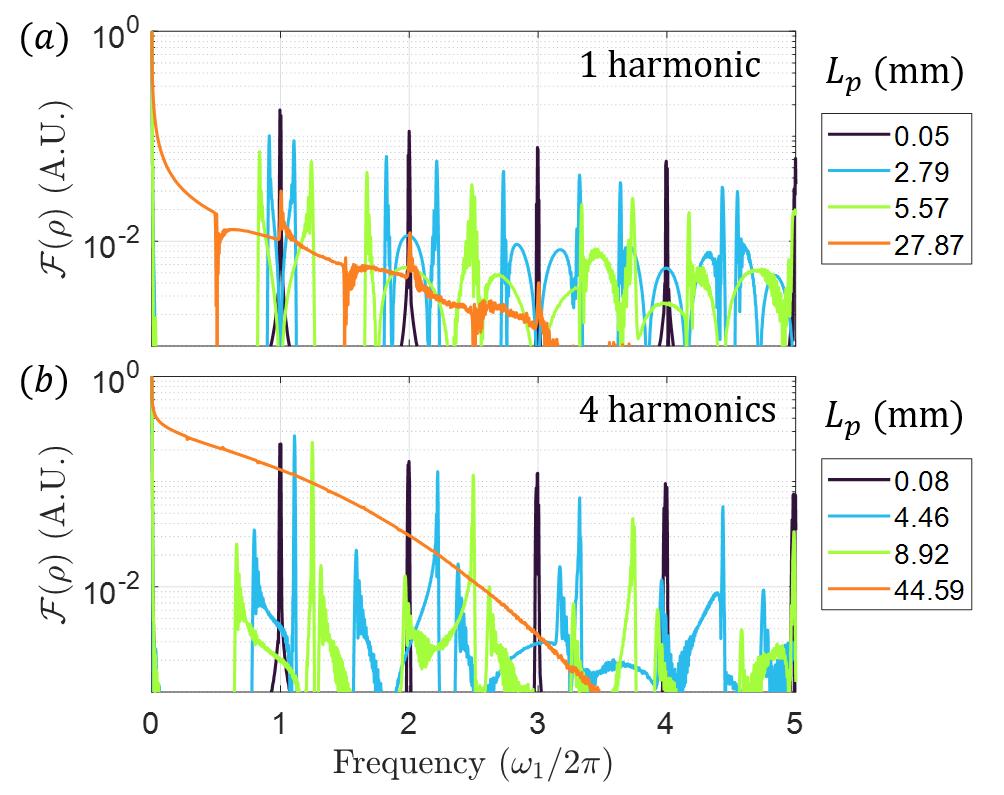}
    \caption{Spectra of electron density at different drift lengths, as indicated by different colors. The optical pulses and electron pulse are the same as for Fig.\ \ref{fig:case_study}. (a) and (b) represent the compression with 1 harmonic (fundamental frequencies) and 4 harmonics respectively.}
    \label{fig:fourier}
\end{figure}

Under the illumination of two optical pulses with central frequencies $\omega_1$ and $\omega_2$, assuming $\omega_1>\omega_2$, the energy modulation becomes $\Delta E(\tau)=\mathcal{E}_1A_1(\tau)\cos(\omega_1 \tau - \phi_1) + \mathcal{E}_2 A_2(\tau)\cos(\omega_2\tau - \phi_2)$.
We assume that the two pulses have the same envelope and that the beat pattern has a node at $\tau=0$, i.e., $\mathcal{E}_1 = \mathcal{E}_2 = \mathcal{E}$, $A_1(\tau) = A_2(\tau) = A(\tau)$, and $\phi_1 -\phi_2 = \pi$. Since $\phi_1 + \phi_2$ has no important influence on compression, we take $\phi_1 = -\phi_2 = \pi/2$ and simplify the energy modulation as
\begin{equation}
    \label{eq:DeltaE_simplify}
    \Delta E(\tau) = 2\mathcal{E}A(\tau)\sin\Big(\frac{\omega_1-\omega_2}{2}\tau\Big) \cos\Big(\frac{\omega_1 + \omega_2}{2} \tau \Big),
\end{equation}
where we call the last term the fast-oscillating part and the rest the slow-varying part ($\Delta\bar{E}(\tau)$). 
We sketch the energy modulation as a function of time $\tau$ near the beat node in Fig.\ \ref{fig:schematic}(b). 
We assume that the electron pulse center passes the modulation stage at $\tau = 0$. The time separation between neighbour beat nodes should be smaller than the driving laser duration ($\Delta_p$) and much larger than the electron pulse duration ($\Delta_e$), such that the electron pulse is almost entirely within a beat node near $\tau = 0$. To satisfy these requirements, we choose $(\omega_1 - \omega_2)/2\pi = 1/4\Delta_e$ and $\Delta_p=20\Delta_e$, but a different parameter choice would not significantly influence the main results. We set $\Delta_e=0.1$ ps  in Fig.\ \ref{fig:schematic} ($\Delta_e=1$ ps in Figs.\ \ref{fig:2freq_2h}-\ref{fig:fourier}), and $\omega_1/2\pi=$ 133.84 THz. 

The compression mechanism is as following.
The slow-varying part in Eq. \ref{eq:DeltaE_simplify} can be linearized near $\tau=0$, 
\begin{equation}
    \label{eq:DeltaE_linearized}
    \Delta\bar{E}(\tau) \approx \alpha \mathcal{E}(\omega_1 - \omega_2)\tau.
\end{equation}
In this case, the scaling factor $\alpha=1$.
With the energy modulation experienced by the electron pulse, as described by Eq.\ (\ref{eq:DeltaE_linearized}), the leading part of the electron pulse ($\tau < 0$) loses energy and the trailing part of the electron pulse ($\tau > 0$) gains energy. 
The electron pulse reaches optimal bunching after a free-space drift length
\begin{equation}
    \label{eq:Lp}
    L_p = \frac{mc^3\beta^3\gamma^3}{\alpha \mathcal{E}(\omega_1 - \omega_2)},
\end{equation}
where we assume the electron pulse has negligible initial energy spread with respect to $\mathcal{E}$, $\beta = v/c$, and $\gamma = 1/\sqrt{1 - \beta^2}$. 
In this ideal limit, the electron pulse with initial duration $\Delta_e$ and energy spread $\delta E_0$ can be compressed to a pulse with duration $\delta E_0/\alpha \mathcal{E}(\omega_1 - \omega_2)$.

In our optical compressor, the initial electron pulse, whose distribution in the energy-time phase space is shown in Fig.\ \ref{fig:schematic}(c), experiences an energy modulation given by Eq.\ (\ref{eq:DeltaE_simplify}). Immediately after the modulation, its phase-space distribution is shown by Fig.\ \ref{fig:schematic}(d). We notice that, at time $\tau_m = 4\pi m/(\omega_1 + \omega_2)$, where $m$ is an integer, the curve $\Delta E(\tau)$ is tangent to the slow-varying part $\Delta\bar{E}(\tau)$. Thus, the clusters of electrons with $\tau$ near $\tau_m$, as indicated by the orange circles in Fig.\ \ref{fig:schematic}(b), experience an energy modulation given by Eq.\ (\ref{eq:DeltaE_linearized}) approximately and can be bunched. After the modulation, the electron pulse undergoes a shear transform in the energy-time phase space when it drifts in the free space, as indicated by the orange arrows in Fig.\ \ref{fig:schematic}(b). Figure \ref{fig:schematic}(e) shows the final phase-space distribution, where the clusters with $\Delta\tau\approx 0$ but different energies form the density peak.

\emph{Improvement with high harmonics --}
The main limitation of this optical electron compressor is that only a small portion of the electrons can be captured in the central density peak after compression. We find that this limitation can be ameliorated if higher harmonics are included in the optical pulses.
To increase the ratio of compressed electrons, we want (1) the slow-varying part $\Delta\bar{E}(\tau)$ of the energy modulation to be closer to a linear function around $\tau=0$ and (2) a larger portion of the total energy modulation $\Delta{E}(\tau)$ to be aligned with $\Delta\bar{E}(\tau)$ such that each cluster can include more electrons. Therefore, with the first $N$ harmonics, i.e. $\omega_1, 2\omega_1, \cdots N\omega_1$, $\omega_2, 2\omega_2, \cdots N\omega_2$, we require that the second to the $2N$-th derivatives of $\Delta E(\tau)$ vanish at $\tau = 0$. 
The energy modulation is thus
\begin{equation}
    \label{eq:DeltaE_Nh}
    \Delta E(\tau) = \mathcal{E}A(\tau)\sum_{l=1,2}\big[(-1)^{l+1}\sum_{n=1}^N a_n \sin(n\omega_l \tau)\big],
\end{equation}
where the coefficients
\begin{equation}
    \label{eq:an}
    a_n = (-1)^{n+1} \frac{(N+1)!(N-1)!}{n(N+n)!(N-n)!}.
\end{equation}
The slow-varying part contributing to the compression is 
\begin{equation}
    \label{eq:DeltabarE_Nh}
    \Delta\bar{E}(\tau) = 2\mathcal{E}A(\tau) \sum_{n=1}^N a_n \sin\big(n \frac{\omega_1 - \omega_2}{2} \tau\big).
\end{equation}

As a demonstration, we include the first 4 harmonics of $\omega_1$ and $\omega_2$. 
There are two important aspects in the energy modulation (Eq.\ (\ref{eq:DeltaE_Nh}) with $N=4$) shown in Fig.\ \ref{fig:2freq_2h}. First, the dashed curve $\Delta\bar{E}(\tau)$ is close to a linear function (Eq.\ (\ref{eq:DeltaE_linearized})) around $\tau=0$, with a slope $\alpha = \sum_{n=1}^N(-1)^{n+1}\frac{(N+1)!(N-1)!}{(N+n)!(N-n)!}$, in a larger time range.
Second, from the zoom-in view (Fig.\ \ref{fig:2freq_2h}(b)), we observe that when $\Delta E(\tau)$ is tangent to $\Delta\bar{ E}(\tau)$ at $\tau_m=4\pi m/(\omega_1 + \omega_2)$, larger portion of electrons experience energy modulation given by the slow-varying part. These two aspects are the two main reasons for the increased ratio of electrons captured in the density peak.
In our numerical example (Fig. \ref{fig:case_study}), this ratio is increased from 3\% with $N=1$ to 21\% with $N=4$.

\emph{Compression results --} We present a demonstration of the optical electron beam compressor with classical particle tracking in the longitudinal phase space. We assume that the electrons are sub-relativistic with initial kinetic energy 57 keV and energy spread 1 eV. The electron pulse has an initial duration of 1 ps with a Gaussian time distribution (Fig.\ \ref{fig:case_study}, orange curve). 
The time-dependent energy modulation is given by Eq.\ (\ref{eq:DeltaE_simplify}) with $\mathcal{E}=400$ eV, $\omega_1/2\pi = 133.84$ THz,  $\omega_2/2\pi=133.59$ THz, and $\Delta_p=20$ ps.
After an optimal free-space drift of 44.6 mm (fine tuned around Eq.\ (\ref{eq:Lp})), the electron density distribution is shown by the blue curve in Fig.\ \ref{fig:case_study}. We find a dominant density peak at the electron pulse center. The compressed pulse has a FWHM duration of 2.3 fs, much smaller than the initial 1 ps pulse duration. 
Nevertheless, not all the electrons can be captured in this density peak. The peak ratio, which is the portion of electrons within the FWHM, is about 3\%. 

The peak ratio improves significantly with the addition of high harmonics. The electron density distribution after the compression with the first 4 harmonics (Eq.\ (\ref{eq:DeltaE_Nh}) with $N=4$) is shown in red in Fig.\ \ref{fig:case_study}. The FWHM is 2.9 fs and the peak ratio is 21\%.

We also study the FWHM of the compressed electron pulse and the peak ratio as a function of the modulation amplitude, as shown by the circles in Fig.\ \ref{fig:fwhm_ratio}. The drift length is fine tuned near the analytical value (Eq.\ (\ref{eq:Lp})) such that the FWHM is minimized. We find that both the FWHM and the peak ratio decrease as the modulation amplitude increases. 
The trend of FWHM is consistent with the theoretical compressed duration $\delta E_0/\mathcal{E}(\omega_1 - \omega_2)$. As a comparison with conventional THz compression, we  study the case that the modulation is the slow-varying $\Delta\bar{E}(\tau)$ (Fig.\ \ref{fig:fwhm_ratio}, dashed curves). The difference in FWHM between optical and THz compression is only about 20\%.
With more harmonics, as shown in Fig.\ \ref{fig:fwhm_ratio}, the peak ratio increases significantly, while the FWHM remains almost unchanged with modulation strength smaller than $\sim 500$ eV and only increases slightly with larger modulation strength when the electron recoil effects are non-negligible. 

\emph{Spectrum of electron density --} We analyze the Fourier components of the electron density distribution as the electron pulse drifts after the energy modulation and find that the dominant frequency varies as it drifts (Fig.\ \ref{fig:fourier}(a)). When interacting with other systems, such electron pulse can provide resonant enhancement with a spatial-dependent resonant frequency.
The physical intuition is that the electrons which pass the modulation stage around time $\tau_m = 4\pi m/(\omega_1 + \omega_2)$, as indicated by the orange circles in Fig.\ \ref{fig:schematic}(b), form a train of periodic clusters. As the electron pulse drifts, the distances between these clusters decrease and the corresponding frequency increases. On the other hand, the electrons passing the modulation stage around time $\tau_m'= 2\pi (2m+1)/(\omega_1 + \omega_2)$ also forms a train of clusters. These clusters separate further as the electrons drift and the corresponding frequency decreases.

The spectra of electron density after modulation by the fundamental frequencies and the first 4 harmonics are shown in Fig.\ \ref{fig:fourier}(a) and Fig.\ \ref{fig:fourier}(b) respectively, where different colors represent different drift distances. 
At a small drift distance (dark blue curve), which corresponds to the micro-bunching distance ($mc^3\beta^3\gamma^3/\alpha\mathcal{E}\omega_1$),
the dominant frequencies are around $M\omega_1$ ($\approx M \omega_2$). When the electron pulse drifts for a distance $L_p$, the dominant frequencies are around
\begin{equation}
    \label{eq:fn}
    f (L_p) = \frac{M}{1\pm \frac{\alpha_{\pm} \mathcal{E}(\omega_1 - \omega_2)L_p}{mc^3\beta^3\gamma^3}}\frac{\omega_1 + \omega_2}{2},
\end{equation}
where $M=1,2,\cdots$ is the order of harmonics, ``$-$'' and ``$+$'' signs are for the frequency increasing and decreasing branches respectively, and $\alpha_{-}=\alpha$, $\alpha_+ = \sum_{n=1}^N\frac{(N+1)!(N-1)!}{(N+n)!(N-n)!}$. 
Equation \ref{eq:fn} is confirmed by the blue and green curves in Fig.\ \ref{fig:fourier} where the drift distances are 0.1 and 0.2 of the optimal compression distances, respectively.
At the optimal compression position, the electron density spectrum is centered at zero frequency, which is shown by the orange curve in Fig.\ \ref{fig:fourier}.
With 4 harmonics, 
comparing Fig.\ \ref{fig:fourier}(b) with Fig.\ \ref{fig:fourier}(a), the amplitudes of the Fourier components increase on the increasing-frequency branch and become larger than those on the decreasing-frequency branch. The large Fourier components make such electron pulses suitable for coherent interactions with classical or quantum systems \cite{gover2020free, karnieli2021coherence, kfir2021optical, zhao2021quantum, ruimy2021toward}. Furthermore, if the optical fields become continuous waves and the electron beam is continuous, such optical compressor can produce a train of density peaks with ps periodicity whose spectrum is a frequency comb.

\emph{Conclusions and Outlooks -- }
In conclusion, we study an optical electron pulse compressor which utilizes the optical beat note to modulate the electron energy. 
We demonstrate fs compression of electron pulses with ps initial duration and 1 eV initial energy spread, achieving a compressed pulse duration comparable to the THz compression. During the free-space drift portion of the compression technique, the dominant Fourier component of the electron density distribution changes. This property can be used to coherently interact with classical or quantum systems, where the resonant frequency is spatially dependent. We also find that by including the higher harmonics in the driving optical pulses, ratio of electrons captured in the density peak can be increased significantly. This compression mechanism can be applied to a wide range of initial electron pulse duration by tuning the optical beat frequency. 
Our classical treatment should be consistent with the quantum description in the region where the electron wave packet duration is smaller than the laser period \cite{zhou2019quantum}. Quantum effects may emerge in the regime where the electron wave packet duration is comparable or longer than the laser period \cite{zhou2019quantum}, or even the beat period. The quantum description of the electron wave packet modulated in such a two-color field may point to new possibilities for electron wave function engineering \cite{barwick2009photon, feist2015quantum, kirchner2014laser, baum2017quantum, zhou2019quantum, reinhardt2020theory, yalunin2021tailored}.  We will investigate such quantum effects in future work.
We believe that our study opens up new opportunities in all-optical control of electrons. 

\emph{Acknowledgements --} We thank members of the ACHIP collaboration for helpful discussions and suggestions. This work is supported by Gordon and Betty Moore Foundation (GBMF4744). Z.\ Zhao and D.\ S.\ Black also acknowledge the Siemann Graduate Fellowship.

\bibliography{dla_thesis_bib.bib}{}
\bibliographystyle{apsrev4-1}  
\end{document}